# Plug-in Electric Vehicle Charging Congestion Analysis Using Taxi Travel Data in the Central Area of Beijing

Huimiao Chen, *Student Member, IEEE*, Hongcai Zhang, *Student Member, IEEE*, Zechun Hu, *Senior Member, IEEE*, Yunyi Liang, Haocheng Luo, Yinhai Wang

*Abstract*— Recharging a plug-in electric vehicle (PEV) is more time-consuming than refueling an internal combustion engine vehicle. As a result, charging stations may face serious congestion problems during peak traffic hours in the near future with the rapid growth of PEV population. Considering that drivers' time costs are usually expensive, charging congestion will be a dominant factor that affect a charging station's quality of service. Hence, it is indispensable to conduct adequate congestion analysis when designing charging stations in order to guarantee acceptable quality of service in the future. This paper proposes a data-driven approach for charging congestion analysis of PEV charging stations. Based on a data-driven PEV charging station planning model, we adopt the queueing theory to model and analyze the charging congestion phenomenon in these planning results. We simulate and analyze the proposed method for charging stations servicing shared-use electric taxis in the central area of Beijing leveraging real-world taxi travel data.

## I. INTRODUCTION

AS a cleaner mode of transport, plug-in electric vehicles (PEVs) have been long considered as a promising tool to combat the energy crisis and climate change. Hence, governments around the world have released extensive incentive policies to popularize them [1].

Different from internal combustion engine vehicles, PEVs need more time to refuel (recharge) so that charging congestion might occur at PEV charging stations. However, this factor is not considered adequately in most literature on PEV charging stations planning [2]-[7]. In the limited number of papers referring to the charging congestion, the overall charging process, including arriving, waiting and charging, is usually modeled by the queueing theory. For example, in [8] and [9], *M/M/s* queueing systems are used, while an *M/G/s/k* queueing system is adopted in [10]. Authors of [11] leverage an *M/M/s/k* queueing model to estimate the probability of electric taxis being charged at their dwell places, but the accurate model is approximated by means of regression and logarithmic transformation.

To the best of our knowledge, there is few paper focusing on the specific and detailed analysis of charging congestion in PEV charging stations. Thus, in this paper, we combine the queueing theory and the real-world taxi travel data to study this problem. The main procedures and contributions of the paper are summarized below.

1) Extract the PEV charging demands from the taxi travel data in the central area of Beijing;
2) Provide a typical median-based location model for PEV charging station planning;
3) Use the queueing theory to model PEV charging congestion, calculate the mean charging waiting time and waiting probability of PEV charging station;
4) Analyze the charging congestion under different charging station planning results and among different charging stations.

The remainder of the paper is organized as follows. Section II describes the taxi travel data and forecasts the PEV charging demands. In Section III, a basic PEV charging station planning method is provided and the corresponding results are shown. In Section IV, the calculation and analysis of charging congestion is elaborately presented. Finally, conclusions are drawn in Section V.

## II. DATA-DRIVEN PEV CHARGING DEMANDS FORECASTING

First, we introduce our data set for this research. The data include 29709 taxis' travel records in the central area, around within the fifth ring, of Beijing from July 1st to July 31st in 2016, which were collected by smart phones or on-board devices. The taxis fleet recorded in the data account for 44% of all the taxis (about 67,000 in total [12]) in Beijing. For each travel record, the information contains the taxi ID, the time and the position (in longitude and latitude). Table I gives a sample of the taxi travel records in the data set.

Based on the data, we further forecast the PEV charging demands. In this research, we assume that ten percent of the taxis in Beijing are replaced by plug-in hybrid electric vehicles, which are still capable of driving by consuming petroleum after the battery power is exhausted or below a threshold. Hence the travel behavior of electric taxis can be supposed to be similar to traditional taxis [13]-[15]. Considering that

This work was supported in part by the National Natural Science Foundation of China under Grant 51477082.

H. Chen, H. Zhang, Z. Hu and H. Luo are with the Department of Electrical Engineering, Tsinghua University, Beijing, 100084, P. R. China (email: chenhm15@mails.tsinghua.edu.cn).
Y. Liang is with Key Laboratory of Road and Traffic Engineering of the Ministry of Education, Tongji University, Shanghai, 201804, P. R. China.
Y. Wang is with the Department of Civil and Environmental Engineering, University of Washington, Seattle, WA, 98195, USA.

recharging battery is more time-consuming than refueling an internal combustion engine vehicle, we regard the dwelling time of at least 30 minutes as the available recharging time windows. In light of the specifics of some plug-in hybrid electric vehicles on the market [16], [17], vehicles' battery capacity and electric range are respectively set as 10 kWh and 50 km. Besides, the rated power of the chargers to be deployed is supposed to be 10 kW. Then, the charging demands for each dwelling of more than 30 minutes can be calculated by the previous vehicle traveled miles. Finally, we can obtain the charging demand point positions and their charging demand weights. Fig. 1 shows the charging demand point distribution of a typical day.

TABLE I. TAXI TRAVEL RECORD SAMPLE

| Taxi ID | Time | Longitude | Latitude |
|---|---|---|---|
| 26491 | 20160704141051 | 116.426285 | 39.921867 |

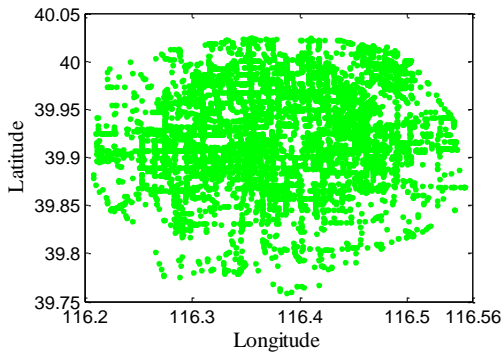

Fig. 1. Charging demand point distribution of a typical day.

### III. DATA-DRIVEN CHARGING STATION PLANNING METHOD

Here, as the basis for the charging congestion analysis in the later section, we provide a typical median-based location model for PEV charging station. In this model, $p$ charging stations are located to minimize the total charging demand weighted distance between charging demand points and the corresponding nearest charging stations. The $p$-median model is formulated as below.

$$\min \sum_{j \in \mathbf{U}} \sum_{i \in \mathbf{V}} D_i L_{ij} A_{ij} \quad (1)$$

subject to:

$$\sum_{j \in \mathbf{U}} A_{ij} = 1, \forall i \in \mathbf{V} \quad (2)$$

$$A_{ij} - B_j \leq 0, \forall i \in \mathbf{V}, j \in \mathbf{U} \quad (3)$$

$$\sum_{j \in \mathbf{U}} B_j = p \quad (4)$$

$$B_j \in \{0,1\}, \forall j \in \mathbf{U} \quad (5)$$

$$A_{ij} \in \{0,1\}, \forall i \in \mathbf{V}, j \in \mathbf{U} \quad (6)$$

In the above model, $D_i$ is charging demands at location $i$; $L_{ij}$ is the trip distance between location $i$ and location $j$; $A_{ij}$ is an assignment variable, which equals 1 if the PEV at location $i$ is assigned to the PEV charging station at location $j$, and 0 otherwise; $B_j$ is a deployment configuration variable, which equals 1 if we site a PEV charging station at location $j$ and 0 otherwise; $p$ is the total number of PEV charging stations to deploy; $\mathbf{U}$ is the set of candidate locations of PEV charging stations and $\mathbf{V}$ is the set of charging demand points. The objective function (1) minimizes the total charging demand weighted distance for PEVs driving to the stations, which describes the convenience of charging services. Constraint (2) ensures that all the charging demands are assigned, while constraint (3) ensures assignments of charging demands to a location with PEV charging station deployed. Constraint (4) shows that the total number of PEV charging stations to be deployed is $p$. Constraints (5) and (6) state that $A_{ij}$ and $B_{ij}$ are binary variables. Constraints (6) can be relaxed to (7) without any sacrifice of optimality, because for any given PEV charging station deployment, charging demands will be assigned to the closest station to achieve the minimum objective.

$$0 \leq A_{ij} \leq 1, \forall i \in \mathbf{V}, j \in \mathbf{U} \quad (7)$$

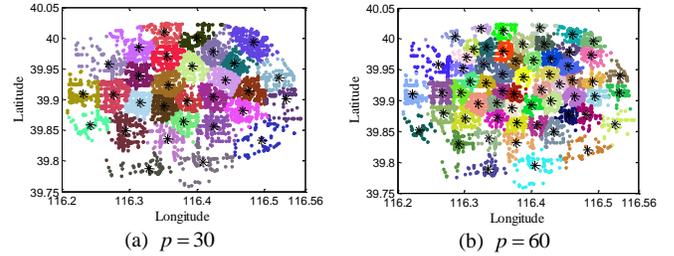

Fig. 2. Results of PEV charging station deployment and charging demand assignment, where the asterisks are the positions of PEV charging stations and the dots of different colors represent the charging demand assignment.

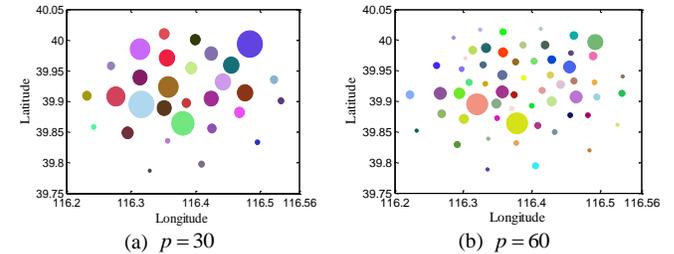

Fig. 3. Comparisons of number of chargers among PEV charging stations, where the radiuses of the circles represent the number of chargers.

The formulated model (1)-(5), (7) is a mix integer linear programming (MILP) model, which can be solved by deterministic branch-and-bound methods. As for the number and the distribution of candidate locations of PEV charging stations, i.e., $\mathbf{U}$, a relatively larger number, e.g., 500, and a uniform distribution are suggested under general scenarios. When $p = 30$, Fig. 2 (a) presents the results of PEV charging station deployment and the corresponding charging demand assignment, and Fig. 3 (a) depicts the comparisons of the number of chargers among the PEV charging stations. The corresponding results for $p = 60$ are given in Figs. 2 (b) and 3 (b).

### IV. CHARGING CONGESTION ANALYSIS USING QUEUEING MODEL

For a PEV charging station, the arrival, waiting and charging of PEVs can be modeled mathematically using queueing theory [8]-[10]. According to [8]-10], the arrival process of PEVs can be considered as a Poisson process, and the service time, i.e., the time to charge each PEV, is supposed to follow the negative exponential distribution. However, in practice, the charging time of PEVs will be affected by various factors, such as the vehicle miles traveled [18], the parking duration [19] and the battery capacity [15]. As a result, the service time is inexplicitly distributed. So, herein, we apply the general distribution to describe the service time of PEVs [10].

Based on the above considerations, the PEV queue in a PEV charging station can be modelled as an *M/G/s/k* queueing system ($k \geq s$), where *M* represents that the time between PEV arrivals to the queue obeys the negative exponential distribution, i.e., the arrival process of PEVs is a Poisson process, *G* represents that the service time of PEVs obeys the general distribution, *s* denotes the number of chargers, and *k* is the total capacity of the queue system, i.e., the summation of the number of chargers of PEV charging stations and the capacity of the waiting spaces. For simplicity, in this paper, the waiting spaces are assumed to be sufficient, i.e., $k = \infty$, and the *M/G/s* queueing system is thereby adopted. Note that the queueing discipline is first come first served (FCFS), i.e., the PEVs are served in the order they arrived in, and the size of calling source, i.e., the population from which the PEVs come, is assumed to be infinite because the total PEV population is large enough so that the arrival rate of PEVs with charging demands will not fluctuate anomalously.

### A. Mean Waiting Time

Armed with the queueing model of a PEV charging station, we are now equipped to consider the calculation of the waiting time of PEV charging. Leveraging the approximation for an *M/G/s* queueing system developed in [20] and [21], we can approximatively compute the mean waiting time of PEVs, expressed as below:

$$W^{M/G/s} = \frac{(1+\xi^2)W^{M/M/s}W^{M/D/s}}{2\xi^2 W^{M/D/s} + (1-\xi^2)W^{M/M/s}} \quad (8)$$

where

$$W^{M/M/s} = \frac{\lambda^s}{(s-1)!(s\mu-\lambda)^2 \mu^{s-1}} \cdot \left[\sum_{z=0}^{s-1} \frac{1}{z!}\left(\frac{\lambda}{\mu}\right)^z + \frac{\lambda^s}{(s-1)!(s\mu-\lambda)\mu^{s-1}}\right]^{-1} \quad (9)$$

$$W^{M/D/s} = \frac{1}{2}\left[1 + H\frac{s\mu-\lambda}{\lambda}\left(1-e^{-\frac{\lambda(s-1)}{H(s\mu-\lambda)(s+1)}}\right)\right]W^{M/M/s} \quad (10)$$

$$H = \frac{s-1}{16s}\left[\left(\frac{10s+8}{2}\right)^{\frac{1}{2}} - 2\right] \quad (11)$$

$$\xi = \sigma\mu \quad (12)$$

In (8)-(12), $W^{M/G/s}$ is the mean waiting time of an *M/G/s* queueing system, and $W^{M/M/s}$ and $W^{M/D/s}$ are respectively the mean waiting times of the corresponding *M/M/s* and *M/D/s* queueing systems; $\lambda$ denotes the arrival rate of PEVs; $\mu$ and $\sigma$ are the reciprocal of the mean charging time of PEVs and the standard deviation of charging time, respectively. Note that (8)-(10) hold when the PEV arrival rate is less than the system transmission capacity, i.e., the utilization factor $\rho = \lambda/s\mu < 1$. Interested readers can refer to [20] and [21] for the details.

While calculating the mean waiting time, the PEV arrival rate $\lambda$ is a constant, i.e., the arrival process of PEVs is a homogeneous Poisson process. But actually, the arrival rate significantly depends on the number of PEVs with charging demand nearby the PEV charging station, which varies with time. Thus, in the real world situation, the PEV arrivals should be equivalent to a non-homogeneous Poisson process with a time-varying $\lambda$, i.e., $\lambda(t)$. In this research, we discretely regard $\lambda$ within an hour as a constant and focus on the mean waiting time in the peak hour.

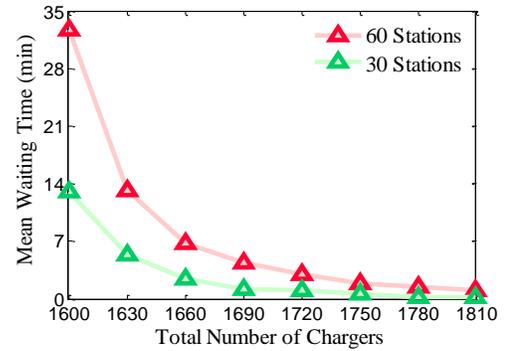

Fig. 4. Mean waiting time curves as the total number of chargers changes.

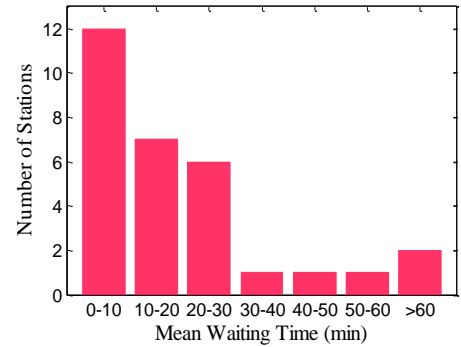

Fig. 5. Distribution of the number of stations with respect to the mean waiting time of stations when the total number of chargers is 1600 and $p = 30$.

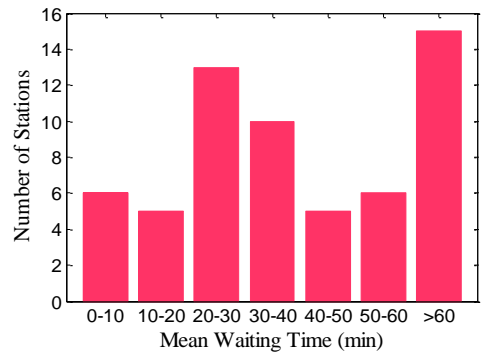

Fig. 6. Distribution of the number of stations with respect to the mean waiting time of stations when the total number of chargers is 1600 and $p=60$.

When $p=30$ and $p=60$, Fig. 4 shows how the mean waiting time changes as the total number of chargers varies. From the figure, it can be seen that 1) more chargers bring less mean waiting time; 2) for a given number of chargers, less number of stations can achieve lees mean waiting time, because centralized chargers are shared by more PEVs and their utilization are higher. Note that a defect of less station is that the distance for a PEV to go to the nearby station becomes longer. For a given number of total chargers (1600 chargers here), the distribution of the number of stations with respect to the mean waiting time of stations when $p=30$ and $p=60$ are respectively presented in Figs. 5 and 6. It can be easily observed that more stations leads to longer mean waiting time when $p=60$. Actually, for all the station, the system transmission capacities, i.e., $\rho$, are nearly the same, that is to say, the proportions of the number of PEVs with charging demands to the number of chargers are almost equal. The reason, why the mean waiting time of different stations differs, is primarily due to the different numbers of PEVs with charging demands, i.e., $s$, for different stations. According to (8)-(12), for a fixed $\rho$, the eventual mean waiting time also depends on $s$ and $\xi$, and the former plays a dominant role.

*B. Waiting Probability*

Let $N$ denote the number of PEVs either waiting or being charged at the station and $P(N)$ denote the probability that $N$ PEVs are at the station. Then $P(N>s)$ is the waiting probability, i.e., the probability that an arriving PEV need to wait. For an *M/M/s* queueing system, it is well known that

$$P^{M/M/s}(N) = \begin{cases} \dfrac{(s\rho)^N}{N!}P_0, & N=0,1,\cdots,s-1 \\ C(1-\rho)\rho^{N-s}, & N \geq s \end{cases} \quad (13)$$

where

$$P_0 = \left[\sum_{z=0}^{s-1}\dfrac{1}{z!}\left(\dfrac{\lambda}{\mu}\right)^z + \dfrac{\lambda^s}{(s-1)!(s\mu-\lambda)\mu^{s-1}}\right]^{-1} \quad (14)$$

and $C$ is the delay probability that

$$C = s\mu(1-\rho)W^{M/M/s} \quad (15)$$

For an *M/G/s* queueing system, geometric approximation is suggestion based on (13) to calculate $P^{M/G/s}(N)$ [22], shown as below.

$$P^{M/G/s}(N) = \begin{cases} \dfrac{(s\rho)^N}{N!}P_0, & N=0,1,\cdots,s-1 \\ C(1-\varsigma)\varsigma^{N-s}, & N \geq s \end{cases} \quad (16)$$

where

$$\varsigma = \dfrac{\rho W^{M/G/s}/W^{M/M/s}}{1-\rho+\rho W^{M/G/s}/W^{M/M/s}} \quad (17)$$

Equation (17) can be derived easily by the Little's formula [23] and $\sum_{N=0}^{\infty}P^{M/G/s}(N)=1$. According to (16), the waiting probability $P^{M/G/s}(N>s)$ can be calculated by (18).

$$P^{M/G/s}(N>s) = 1 - \sum_{N=1}^{s-1}\dfrac{(s\rho)^N}{N!}P_0 - C(1-\varsigma) \quad (18)$$

When $p=30$ and $p=60$, we plot the waiting probability variation curves with respect to the total number of chargers, as shown in Fig. 7. The results are similar to those of the mean waiting time, and can be interpreted by the same reason in Subsection IV.A. Also, letting the number of total chargers be 1600, the distribution of the number of stations with respect to the waiting probability of stations when $p=30$ and $p=60$ are respectively presented in Figs. 8 and 9. It can be observed that compared with the results in Fig. 10, in Fig. 11, the stations with relatively larger waiting probability accounts for a higher percentage. The analysis for the different waiting probabilities among the different stations can refer to that for mean waiting time.

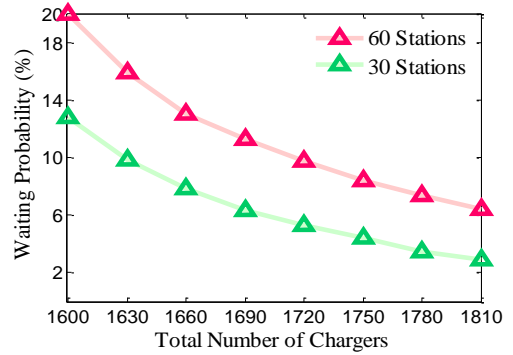

Fig. 7. Waiting probability curves as the total number of chargers changes.

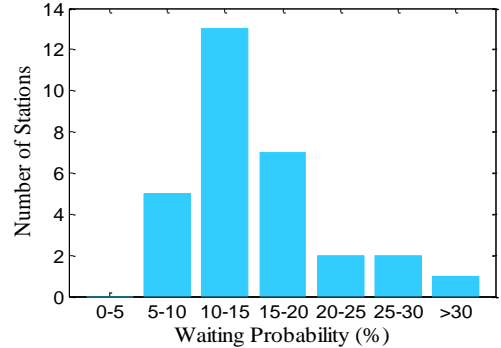

Fig. 8. Distribution of the number of stations with respect to the waiting probability of stations when the total number of chargers is 1600 and $p=30$.

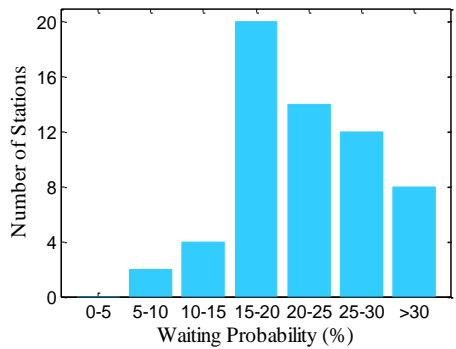

Fig. 9. Distribution of the number of stations with respect to the waiting probability of stations when the total number of chargers is 1600 and $p=60$.

*C. Mean Driving Time to the Station*

In Subsection IV-A, we can see that for a fixed charger number, less stations can achieve less waiting time. However, less stations in a certain area causes more inconvenience for PEV users to recharge their cars. Therefore, there is a tradeoff between waiting time and driving time to the station. We calculate the mean driving time to the station when the charging station number is 30 and 60, respectively. The corresponding results are 2.3 minutes and 1.5 minutes. Combining Figure 6, it can be observed that as the total charger number increases, the difference of mean waiting time between 60 stations scenario and 30 stations scenario tends to be closing. Thus, when the total charger number is limited, for example, less than 1660, it is better to build 30 stations rather than 60 stations. On the contrary, more stations are preferred, when the investment for charging infrastructure is sufficient to install more chargers.

## V. Conclusions and Future Work

In this paper, we combine the queueing theory and real-world taxi travel data in the central area of Beijing to analyze the charging congestion of PEV charging stations. The results show that 1) the mean waiting time and the waiting probability decrease as the total number of the chargers of all the PEV charging stations increases; 2) even though the same number of chargers are deployed, different number of charging stations significantly affects the mean waiting time and the waiting probability of PEVs; 3) in charging stations with the almost same proportion of the number of PEVs to the number of chargers, the difference of the number of chargers may bring distinctly different mean waiting time and waiting probabilities; 4) if the investment for charging infrastructure is sufficient, i.e., more charger can be installed, chargers should be located dispersedly; otherwise, more centralized deployment is better.

In future work, we plan to apply the charging congestion analysis into the PEV charging station planning.